\documentstyle[12pt]{article}
\def\beq{\begin{equation}}
\def\eeq{\end{equation}}
\def\bea{\begin{eqnarray}}
\def\eea{\end{eqnarray}}
\def\bq{\begin{quote}}
\def\eq{\end{quote}}

\begin{document}
\pagestyle{empty}
\begin{flushright}
{ROME prep.1089/95 \\
 hep-th/9503013}
\end{flushright}
\vspace*{5mm}
\begin{center}
{\bf TWO DIMENSIONAL ${\bf QCD}$ AND ABELIAN BOSONIZATION }
\\
\vspace*{1cm}
{\bf M. Bochicchio} \\
\vspace*{0.5cm}
INFN Sezione di Roma \\
Dipartimento di Fisica, Universita' di Roma `La Sapienza' \\
Piazzale Aldo Moro 2 , 00185 Roma  \\
\vspace*{2cm}
{\bf ABSTRACT  } \\
\end{center}
\vspace*{5mm}
\noindent
A bosonized action, that reproduces the structure of the 't Hooft equation
for $QCD_2$ in the large-$N$ limit, up to regularization dependent
terms, is derived.
\vspace*{4cm}
\begin{flushleft}
May 1995
\end{flushleft}
\phantom{ }
\vfill
\eject

\setcounter{page}{1}
\pagestyle{plain}

\section{}

The aim of this paper is to show that the effective action for mesons in
$QCD_2$ is a local one and can be computed by means of abelian bosonization.
As a partial check I also show that the structure of the various terms
that appear in the 't Hooft equation \cite{O}
is reproduced, up to regularization dependent terms, when quantum fluctuations
are computed for this action in the
large-$N$ limit. I leave for a forthcoming paper the check that also the
coefficients of each term agree with the 't Hooft equation .\\
To be precise, in this paper I consider a slightly more general version of
$QCD_2$ which is referred to as the generalized 't Hooft model and has been
introduced in the literature in ref. \cite{Dou}.
In the case of the (generalized) Schwinger model,
the authors of ref. \cite{Dou} have shown how standard bosonization
gives an effective action of mesons interacting with a Landau-Ginzburg
potential.\\
In the present paper I show how to get the meson effective action
in the non-abelian case, starting with the Lagrangian of the generalized 't
Hooft model:
\bea
L=\frac{N}{8\pi} Tr(E\epsilon^{\mu \nu}F_{\mu \nu})-\frac{N}{4\pi}g^2\sum^
{\infty}_{2}
f_{n}TrE^n+\bar\psi(iD_{\mu}\gamma^{\mu}-m)\psi
\eea
This Lagrangian reduces to $QCD_2$ with one flavour for
 $f_2=\frac{1}{8\pi}$ and all other $f_n$ vanishing.
In the $U(1)$ case, the meson effective action follows immediately from
standard bosonization of the fermionic part of the Lagrangian plus
axial gauge-fixing of the gauge field \cite{Dou}. \\
In the non-abelian case, two more crucial steps are necessary for
being able to perform abelian bosonization. The $U(N)$ and $SU(N)$ case can be
considered at the same time. In the following formulae we can go from $U(N)$ to
$SU(N)$ simply imposing the traceless condition on all the diagonal
Lie-algebra generators.
First, it is convenient to fix the gauge partially, reducing the $U(N)$
gauge symmetry to a diagonal $U(1)^{N}$, setting equal to zero the charged
components of $E$:
\bea
E^{ch}=0
\eea
After doing so, the integration over the charged gluons gives the effective
Lagrangian:
\bea
L&=&\frac{N}{8\pi}\sum_k E_k\epsilon^{\mu \nu} (\partial_{\mu} A^k_{\nu}-
\partial_{\nu} A^k_{\mu})-
\frac{N}{4\pi}g^2\sum^{\infty}_{2}f_{n}\sum_k E_k^n +  \nonumber \\
&& +\sum_k \bar\psi^k[(i\partial_{\mu}-A^k_{\mu})\gamma^{\mu}-m]\psi^k
-\frac{4\pi i}{N}
\sum_{k \ne j} \frac{\bar\psi^k \gamma^1 \psi^j \bar\psi^j \gamma^0 \psi^k}
{E_k-E_j}
\eea
where the sum on the indices $k$ and $j$ now runs only over the neutral
diagonal components of the $E$-field.
Remarkably the effective action in Eq.(3) contains a local four fermion
interaction
of the charged fermionic currents. \\
This effective Lagrangian has been
obtained integrating over all paths the exponential of $i$ times the action
associated to the Minkowski
Lagrangian of Eq.(1). I have choosen to perform Minkowski path integrals
instead of Euclidean ones only for convenience, in order to use standard
bosonization formulae. I have not included in the Minkowski effective action
the Faddeev-Popov determinant, that comes from the gauge-fixing, and the
determinant
that comes from the integration over the charged gluon fields \cite{Blau}.
They will be included only at the end of the computation, after analytical
continuation from the Minkowski to the Euclidean region. In any case, their
contribution to the Euclidean Lagrangian (defined in such a way that the
exponential of minus the Euclidean action enters the path integral) is given
by:

\bea
    - \frac{1}{4\pi}\sum_{k>j} R \log(E_k-E_j)^2
\eea
where $R$ is the Riemannian curvature of the surface on which the theory lives.
The second step consists in a Fierz rearranging of the four fermion term,
that produces a new four fermion interaction involving neutral chiral currents:

\bea
L&=&\frac{N}{8\pi}\sum_k E_k \epsilon^{\mu \nu} (\partial_{\mu} A^k_{\nu}-
\partial_{\nu} A^k_{\mu})-
\frac{N}{4\pi}g^2\sum^{\infty}_{2}f_{n}\sum_k E_k^n+  \nonumber \\
&& +\sum_k \bar\psi^k[(i\partial_{\mu}-A^k_{\mu})\gamma^{\mu}-m]\psi^k+
 \nonumber \\
&& +\frac{2\pi i}{N}
\sum_{k \ne j} \frac{\bar\psi^k (1+\gamma^5) \psi^k \bar\psi^j (1-\gamma^5)
\psi^j}
{E_k-E_j}
\eea

At this point a standard trick, used for the Thirring model, allows us to write
the four fermion interaction as the Gaussian integral of a linear one.
Being neutral, each current can be bosonized separately by standard abelian
bosonization \cite{bos}, giving the effective Lagrangian:

\bea
L&=&\frac{N}{8\pi}\sum_k ( E_k-\frac{2}{N}\phi_k ) \epsilon^{\mu \nu}
(\partial_{\mu} A^k_{\nu}-\partial_{\nu} A^k_{\mu})-
\frac{N}{4\pi}g^2\sum^{\infty}_{2}f_{n}\sum_k E_k^n +
    \nonumber \\
&& +\sum_k \frac{1}{8\pi} \partial^{\mu} \phi_k \partial_{\mu} \phi_k
 - m c \mu \cos \phi_k
 -\frac{4\pi}{N} (c \mu)^2
\sum_{k>j} \frac{\sin(\phi_k-\phi_j)}
{E_k-E_j}
\nonumber \\
\eea

where the constants $c$ and the scale $\mu$ are regularization dependent.
After continuing to the Euclidean region and adding the ghost and gluon
contribution of Eq.(4), the Euclidean effective action becomes:

\bea
L_{E}&=&- i \frac{N}{8\pi}\sum_k ( E_k-\frac{2}{N}\phi_k ) \epsilon^{\mu \nu}
(\partial_{\mu} A^k_{\nu}-\partial_{\nu} A^k_{\mu})+
\frac{N}{4\pi}g^2\sum^{\infty}_{2}f_{n}\sum_k E_k^n + \nonumber \\
&& - \frac{1}{4\pi}\sum_{k>j} R \log(E_k-E_j)^2 +  \nonumber \\
&& +\sum_k \frac{1}{8\pi} \partial^{\mu} \phi_k \partial_{\mu} \phi_k
 + m  c \mu \cos \phi_k
 +
\frac{4\pi}{N} (c \mu)^2
\sum_{k>j} \frac{\sin(\phi_k-\phi_j)}
{E_k-E_j}
\nonumber \\
\eea

As a last step, integrating away, in an axial gauge, the neutral $U(1)$ gauge
fields that
enter the first order Lagrangian as Lagrange multipliers,
gives the constraint:

\bea
 E_k=\frac{2}{N}\phi_k + \theta_k
\eea

This constraint identifies the electric field with the bosonized fermion field,
up to a rescaling and a shift by a $\theta$ angle which appears also in the
Schwinger model \cite{Dou}. With the $\theta$ set to zero,
the final result for the (Euclidean) meson effective action is:

\bea
L_{E}&=&
\frac{N}{4\pi}g^2\sum^{\infty}_{2} f_{n} \sum_k (\frac{2 \phi_k} {\sqrt N })^n-
\frac{1}{4\pi}\sum_{k>j} R \log(\phi_k-\phi_j)^2 + \nonumber \\
&& +\sum_k \frac{N}{8\pi} \partial^{\mu} \phi_k \partial_{\mu} \phi_k
 + c \mu m \cos(\sqrt{N}\phi_k) + \nonumber \\
&& +\frac{2\pi}{\sqrt N} (c \mu)^2
\sum_{k>j} \frac{\sin(\sqrt N \phi_k-\sqrt N \phi_j)}
{\phi_k-\phi_j}
\nonumber \\
\eea

where I have rescaled the $\phi_k$ fields by a factor of $\sqrt{N}$. \\
A few comments are in order.\\
At the leading $1/N$ order, the Lagrangian of Eq.(7)
reproduces the topological
Lagrangian of $YM_2$ \cite{W}, if configurations with non-trivial $U(1)$
magnetic
charge are included \cite{Blau}.
In this case, indeed, the $E_k$ fields become integral
valued in the dual of the lattice of the magnetic charges
up to a constant factor. \\
If the integral nature of $E_k$ is ignored, in the large-$N$ limit the
structure of the 't Hooft equation is reproduced, up to the regularization
dependent terms, at least for
the massless case. Indeed the 't Hooft equation:

\bea
p^2 \chi(x)&=&M^2[\frac{1}{x}+\frac{1}{1-x}] \chi(x) - \frac{g^2}{\pi}
P\int^{1}_{0} \frac{\chi(y)}{(x-y)^2} \nonumber \\
M^2 &=& m^2-\frac{g^2}{\pi}
\eea

contains two terms, the kinetic one and the potential one, that can be easily
traced back to the kinetic term for $\phi_k$ and the second derivative of the
logarithm of the Vandermonde determinant in Eq.(9). More care and further study
is necessary to trace
back the term proportional to $m^2$ in the 't Hooft equation to the
regularization dependent terms in Eq.(9). They may well contribute the
necessary
quadratic term in the 't Hooft equation. However, since their coefficients are
regularization dependent, it is necessary to find a way to fix them
unambiguously. Since the regularization implicit in the bosonization need not
preserve chiral symmetry, there should possibly be a mixing between them.
The mixing coefficients should be determined using chiral Ward identities,
in the spirit of \cite{Boc}.
Notice that some care about signs has been
necessary for comparing the 't Hooft equation that is defined in Minkowski
space, with the Lagrangian of Eq.(9) that is the Euclidean one.
However, for a complete comparison of the structure and the coefficients
of the bosonized effective action with the 't Hooft equation, it is necessary
to
compute carefully the
quadratic fluctuations around the master field for this action. This
necessarily involves
the density of the eigenvalues and it is left for further investigation.\\
The integral nature of $E_k$ allows probably a stringy interpretation of
$QCD_2$
analogous to $YM_2$ \cite{Gross}, in which this time the $QCD$ string
propagates real
degrees of freedom and the partition function does not count only covering
maps of Riemann surfaces. \\
It is also clear that the condensation of the eigenvalues, that occurs at
strong coupling in $YM_2$ \cite{Ka}, modifies the 't Hooft equation,
in the strong coupling
limit, via the different density of eigenvalues.
I plan to consider these matters in more detail in a forthcoming
paper.

\section{Acknowledgements}

The author wishes to thank E. Abdalla and C. Abdalla for mentioning,
while this paper was typewritten,
ref. \cite{S} and ref. \cite{B}, where the four fermion interaction term in the
effective action was already
obtained. However,
the complete meson effective action in the abelian bosonized form and the
link with the 't Hooft equation seem to be new in the literature.
It should also be mentioned that in ref. \cite{A} an alternative approach based
on non-abelian bosonization is pursued; it would be interesting to find
out a relation with the abelian bosonization.\\
In addition, in ref. \cite{Di}, the 't Hooft equation has been obtained using
functional techniques in the light-cone gauge.

\end{document}